\documentclass[aps,amssymb,10pt,showpacs,letterpaper,twocolumn]{revtex4}
\usepackage{graphicx}
\usepackage{amsmath}
\usepackage{epsfig}
\usepackage{bm}

\newcommand{\be}{\begin{equation}}
\newcommand{\ee}{\end{equation}}
\newcommand{\beq}{\begin{eqnarray}}
\newcommand{\eeq}{\end{eqnarray}}

\begin{document}

\title{Massive neutral particles on heterotic string theory.}

\author{ Marco Olivares }
\email{marco.olivaresrubilar@gmail.com}
\affiliation{\it Instituto de F\'{\i}sica, Pontificia Universidad de
Cat\'{o}lica de Valpara\'{\i}so, Av. Universidad 330, Curauma,
Valpara\'{\i}so, Chile.}

\author{ J. R. Villanueva }

 \email{jose.villanuevalob@uv.cl}
\affiliation{Instituto de F\'{\i}sica y Astronom\'ia,
  Universidad de Valpara\'iso,
  Gran Breta\~na 1111, Playa Ancha, Valpara\'iso, Chile,}
\affiliation{Centro de Astrof\'isica de Valpara\'iso, Gran Breta\~na 1111, Playa Ancha,
Valpara\'{\i}so, Chile.}

\date{\today}

\begin{abstract}
The motion of massive particles in the background of a
charged black hole in heterotic string theory, which is characterized
by a parameter $\alpha$, is studied in detail
in this paper. Since it is possible to write this space-time
in the Einstein frame, we perform a quantitative analysis of the
time-like geodesics by means of the standard Lagrange procedure.
Thus, we obtain and solve a set of differential equations and then we
describe the orbits in terms of the elliptic
$\wp$-Weierstra{\ss} function.
Also, by making an elementary derivation developed by Cornbleet
(Am. J. Phys. \textbf{61} 7, (1993) 650 - 651)
we obtain the correction to the angle of advance of perihelion to
first order in $\alpha$,  and thus, by comparing with
Mercury's data we give an estimation for the value of
this parameter, which yields an {\it heterotic solar charge} 
$Q_{\odot}\simeq 0.728\,[\textrm{Km}]= 0.493\, M_{\odot}$.
Therefore, in addition  to
the study on null geodesics performed by Fernando (Phys. Rev. D {\bf 85}, (2012)
024033), this work completes the geodesic structure
for this class of space-time.
\end{abstract}


%\keywords{Black Holes;  Elliptic Functions; String Theory.}

\maketitle

\tableofcontents

%%%%%%%%%%%%%%%%%%%%%%%%%

\section{Introduction}

%%%%%%%%%%%%%%%%%%%%%%%%%%%%%%%%%%%%%%%%%%%%%%%%%%%%%%%%%%%%%%%%%%%%%%%%%%%%%%%%%%%%%%%%%%%%%%%%%%%
It is well known that an effective field theory describing string
theory contains black hole solutions which have some properties
that appear different from its counterpart of the standard
General Relativity \cite{gibbons88,garfinkle91}.
In this paper we will try to test those differences between both space-times
by solving the complete time-like geodesics.

The starting point for studying this solutions is the
effective action in heterotic string theory in four dimensions \cite{sen92}
\begin{eqnarray}
  S&=&\frac{1}{16\pi}\int d^4x\sqrt{-g} \left[R-\frac{1}{12}e^{-4\Phi}H_{\mu \nu \lambda}H^{\mu \nu \lambda}+\right. \nonumber\\
  &&\left.-2(\nabla \Phi)^2-e^{-2\Phi}F_{\mu\nu}F^{\mu\nu}\right],
  \label{a0}
\end{eqnarray}
where $\Phi$ is the dilaton field, $R$ is the scalar curvature,
$F_{\mu\nu}=\partial_{\mu} A_{\nu}-\partial_{\nu} A_{\mu}$ is Maxwell's field strength associated with
an $U(1)$ subgroup of $E_8\times E_8$ or Spin(32)/$\mathbb{Z}_2$, and
\begin{equation}\label{a01}
  H_{\mu \nu \lambda}=\partial_{\mu} B_{\nu\lambda}+\partial_{\nu} B_{\lambda \mu}+\partial_{\lambda} B_{\mu\nu}-(\Omega_3(A))_{\mu \nu \lambda},
\end{equation}
where $B_{\mu\nu}$ is the antisymmetric tensor gauge field, and
\begin{equation}\label{a02}
  (\Omega_3(A))_{\mu \nu \lambda}=\frac{1}{4}(A_{\mu}F_{\nu\lambda}+A_{\nu}F_{\lambda\mu}+A_{\lambda}F_{\mu\nu}),
\end{equation}
is the gauge Chern--Simons term. In this contribution we are interested in the
situation when the fields $H_{\mu \nu \lambda}$ and $B_{\mu\nu}$ take the zero value, in which case
action (\ref{a0}) reads
%%%ACTION
\begin{equation}\label{a1}
  S=\frac{1}{16\pi}\int d^4x\sqrt{-g} \left[R-2(\nabla \Phi)^2-e^{-2\Phi}F_{\mu\nu}F^{\mu\nu}\right],
\end{equation}
which leads to the field equations
\begin{eqnarray}
 \label{a001}
  \nabla_{\mu}(e^{-2\Phi}F^{\mu\nu}) &=& 0, \\ \label{a002}
  \nabla^2\Phi+\frac{1}{2}e^{-2\Phi}F^2 &=& 0,
\end{eqnarray}
and
\begin{equation}\label{a003}
  R_{\mu\nu}=-2\nabla_{\mu}\Phi\,\nabla_{\nu}\Phi-2e^{-2\Phi}F_{\mu\lambda}F_{\nu}^{\lambda}
  +\frac{1}{2}g_{\mu\nu}e^{-2\Phi}F^2.
\end{equation}
Solutions to Eqs. (\ref{a001}, \ref{a002}, \ref{a003}) for a static charged black hole
were found by Gibbons and Maeda \cite{gibbons88}, and independently by
Garfinkle, Horowitz and Strominger \cite{garfinkle91}. So, we will refer
to this black hole as the Gibbons--Maeda--Garfinkle--Horowitz--Strominger
(GMGHS) black hole. Thereafter, numerous studies have been performed in the
context of the heterotic string theory.
For example, the solution described a black hole in four dimensions carrying
mass, charge and angular momentum has been found in \cite{sen92}, where also
the extremal limit  of this solution is discussed. The same author constructed
the general electrically charged, rotating black hole solution in the heterotic string theory compactified
on a six-dimensional torus \cite{sen95}. Also, Hassan and Sen \cite{hassan} have shown
that given a classical solution of the heterotic string theory which is independent of a number $d$
of the space-times coordinates, and for which the background gauge field lies in a subgroup
that commutes with $p$ of the $U(1)$ generators of the gauge group, one can
generate other classical solutions by applying an $O(d-1, 1)\otimes O(d+p-1, 1)$
transformations on the original solution. Thus, using this method,
the authors constructed black string solutions in six dimensions carrying electric charge, and both,
electric and magnetic type antisymmetric tensor gauge-field charge.
On the other hand, the introduction of the basic aspects of string solitons,
duality and black holes in string theory can be found in \cite{khuri}, whereas the static solutions
of electrically and magnetically charged dilaton black holes with the topology
of $R^2\otimes S^{n-2}$, $R^2\otimes S^1\otimes S^{n-3}$ and
$R^2\otimes R^1\otimes S^{n-3}$ constructed from the dilaton gravity theory
with cosmological constant was obtained in \cite{gao05}.

In general relativity, the motion of particles in the background
of charged black holes has been studied in various papers.
For example, the motion of particles in higher
dimensional charged static spherically symmetric space-times was presented in \cite{hackmann}.
In four dimensions, the properties of the Reissner--Nordstr\"{o}m space-time with non-zero
cosmological constant was performed in \cite{Hledik}.
Also, the motion of massless
particles on this background with negative cosmological constant can be found in \cite{cosv}, whereas the motion
of charged particles in the same space-time was made by Olivares et al. \cite{monin}.
Stuchl\'ik and Calvani \cite{calvani} have studied the radial motion and perform a
qualitative analysis of the allowed orbits for photons on the charged black hole with $\Lambda>0$,
whereas Pugliese et al. \cite{pugliese} have studied the circular motion of neutral particles
on the Reissner--Nordstr\"{o}m space-time without cosmological constant.

The study of geodesics of charged black holes in string theory
is an important area of research, this is because where the
gravity meets all other fundamental forces in nature and the
classical equation of motion takes the form of Einstein
equations plus Planck scale correction terms.
A primary work by considering the circular motion and the scattering problem
of charged particles by a charged dilatonic black hole with arbitrary coupling
constant $a$ was realized by Maki and Shiraishi \cite{maki94},
however there is no detailed discussion of effective potential
that accounts for all orbits allowed.
In this work we are interested in investigating the time-like geodesics
around a GMGHS black hole which corresponds to the case $a=1$,
since once this task is performed,
geodesic structure for this space-time is completed,
because the null geodesics were resolved fully
by Fernando \cite{fernando}, while the study
of the gravitational lensing was performed in \cite{bhadra}.
Also, the geodesic motion of neutral
test particles for equatorial time-like circular geodesic and
null circular geodesic, both extremal and non-extremal
case of charged black hole in string theory was studied
in \cite{pradhan1}, whereas the geodesic motion in the multiply
warped product space-time near the hypersurfaces in the interior
of the event horizon can be found in \cite{choi}.

This paper is organized as follows: in section \ref{STL} we
presents the procedure to obtain the motion equations of
massive particles in the GMGHS black hole background,
and then we solve these equations to describe the allow orbits
analytically. In section \ref{peri} we apply an elementary derivation
to evaluate the perihelion precession in this space-time.
Finally, in section \ref{summ} we conclude with some comments
and final remarks.

\section{Time-like geodesics}
\label{STL}
\begin{figure}[!h]
 \begin{center}
  \includegraphics[width=83mm]{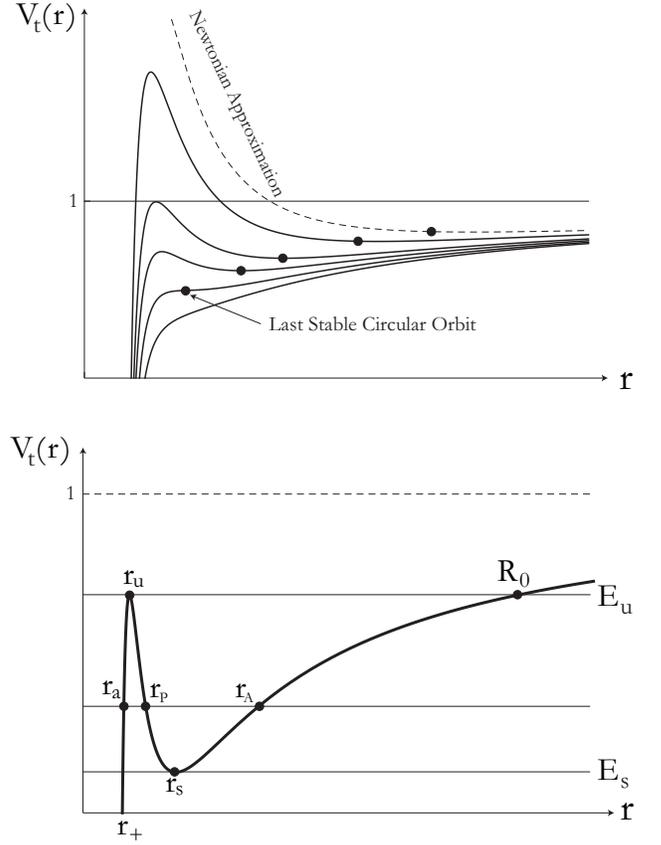}
 \end{center}
 \caption{Top panel: the effective potential for particles with non-vanished angular
 momentum. When the angular momentum takes the value $ L_{lsco} $, then the maximum value of the
 potential coincides with the minimum in a single point (point of inflection), where the
 black hole's last stable circular orbit (lsco) occurs.
Bottom panel: the effective potential for higher values of the critical
angular momentum allows the existence of orbits confined
between two turning points $r_P <r <r_A$, also for $r_S$
stable circular orbit, and an unstable one, $r_u$.}
 \label{f1}
\end{figure}

With the aim to study the motion of massive neutral particles around the GMGHS
black hole, we first derive the geodesic equations  following the same approach
given in \cite{chandra,COV,germancito,weyl,cov21,vv,valeria08,evita0,evita,betti,chen,zhou,sultana,halisoy}.
Thus, we can write the GMGHS metric in the Einstein frame as \cite{fernando}
\begin{equation}\label{tl1}
  ds^2=-\mathcal{F}\, dt^2+\frac{dr^2}{\mathcal{F}}+\mathcal{R}^2\,(d\theta^2+\sin^2\theta\,d\phi^2),
\end{equation}
where the radial function, $\mathcal{R}=\mathcal{R}(r)$, is given by
\begin{equation}\label{tl2}
  \mathcal{R}=\sqrt{r\left(r-\frac{Q^2}{M}\right)}=
  \sqrt{r(r-\alpha)}, \qquad \alpha \equiv \frac{Q^2}{M},
\end{equation}
$M$ is the ADM mass and $Q$ is the electric charge of the GMGHS black hole,
and $\mathcal{F}=\mathcal{F}(r)$ is the well known lapse function of the Schwarzschild black hole,
\begin{equation}\label{tl3}
  \mathcal{F}=1-\frac{2M}{r}= 1-\frac{r_+}{r}, \qquad r_+=2M.
\end{equation}
Also, the coordinates in (\ref{tl1}) are defined in the ranges
$0<r<\infty$, $-\infty<t<\infty$, $0\leq \theta <\pi$, and
$0\leq \phi <2\pi$. So, the normalized Lagrangian associated to the metric
(\ref{tl1}) results:
\begin{equation}
  2\mathcal{L}=- \mathcal{F}\,\dot{t}^2+
  \frac{\dot{r}^2}{\mathcal{F}}+\mathcal{R}^2\,(\dot{\theta}^2+\sin^2\theta\,\dot{\phi}^2)
  =-1 ,\label{tl4}
\end{equation}
where $\dot{a}=da/d\tau$, and $\tau$ is an affine parameter along the geodesic that
we choose as the proper time. Since the Lagrangian (\ref{tl4}) is
independent of the cyclic coordinates ($t, \phi$), then their
conjugate momenta ($\Pi_t, \Pi_{\phi}$) are conserved and are given by
\begin{equation}\label{tl5}
  %-\left(1-\frac{r_+}{r}\right)
  \mathcal{F}\,\dot{t}=-\sqrt{E},
\end{equation}
and
\begin{equation}\label{tl6}
  \mathcal{R}^2 \dot{\phi}=L,
\end{equation}
in the invariant plane $\theta=\pi/2$. So, inserting
Eqs. (\ref{tl5}) and (\ref{tl6}) into Eq. (\ref{tl4}),
we obtain
\begin{equation}\label{tl7}
  \left(\frac{dr}{d\tau}\right)^2=E-V_{t},
\end{equation}
where $V_{t}=V_{t}(r)$ is the effective potential
given by
\begin{equation}\label{tl8}
  V_{t}=\mathcal{F}\,\left(1+\frac{L^2}{\mathcal{R}^2}\right),
\end{equation}
which is showed in Fig. \ref{f1}. From this, we can see that confined orbit can exist
depending on the value of the constant of motion $L$.
Let us start by examining the extreme
condition $dV_t/dr=0$, which leads to the cubic equation
\begin{equation}\label{t19}
r^3- 2L_{\alpha}r^2+
\left( \alpha L_{\alpha}+3 L^2 \right)r-2 \alpha L^2=0,
\end{equation}
where
\begin{equation}\label{tl900}
  L_{\alpha}=\frac{L^2}{r_+}+\alpha.
\end{equation}
So, by making the following identification:
\begin{equation}\label{t190}
  r_{\sigma}={2 L_{\alpha} \over 3},\, \, \,
  R_{\sigma}=\sqrt{{\chi_2 \over 3}},\,\, \, \theta_{\sigma}={1 \over 3} \arccos \sqrt{{27\chi_{3}^2 \over \chi_{2}^{3}}},
\end{equation}
with
\begin{eqnarray}
  \chi_2 &=&4\left[\frac{4}{3}L_{\alpha}^2-(\alpha\,L_{\alpha}+3L^2)\right],\nonumber\\
\chi_3 &=&4\left[\frac{16}{27}L_{\alpha}^3-\frac{2}{3}L_{\alpha}(\alpha\,L_{\alpha}+3L^2)
+2 \alpha L^2\right] ,\label{t192}
\end{eqnarray}
we can write the nth solution of Eq. (\ref{t19}) as a function of the angular momentum,
resulting in
\begin{equation}\label{t191}
r_{ext}^{(n)}(L)=r_{\sigma}
+R_{\sigma}
\cos \left(\theta_{\sigma}+
{2 n \pi  \over 3}\right),\qquad (n=0, 1, 2),
\end{equation}
so, the stable circular orbit is obtained by setting $n=0$ ($r_{ext}^{(0)}(L)\equiv r_s$),
while the unstable circular orbit is obtained by setting $n = 2$ ($r_{ext}^{(2)}(L)\equiv r_u$).
Notice that $r_s \geq r_u$, and  the existence of circular orbits are linked to
the value of the angular momentum. Here the equality $r_s = r_u\equiv r_l$
means that there is a critical value for $L$, say $L_{lsco}$,
as is referred to in the top panel of Fig. \ref{f1}. So, by making
the following identifications:
\begin{eqnarray}
  &&\ell_{L}^2={r_+(3 r_+ -\alpha)^2\over 9 r_+ -\alpha} ,\qquad
  \Gamma_{L}^2=\sqrt{{\lambda_2 \over 3}}\nonumber\\
  && \theta_L={1 \over 3} \arccos \sqrt{{27 \lambda_{3}^2 \over \chi_{2}^{3}}},
  \label{t192.0}
\end{eqnarray}
we can write this critical value as
\begin{equation}\label{t192}
L_{lsco}^{2}= \ell_{L}^2+\Gamma_{L}^2
\cos \theta_L,
\end{equation}
where
\begin{eqnarray}
  \lambda_2 &=& 324{ r_{+} \mathcal{R}^2_+\,\ell_L^4\over (3 r_+ -\alpha)^3} , \\
  \lambda_3 &=& 108\left[{r_{+}^{4} \mathcal{R}^2_+\,(54 r_{+}^{3}-54\alpha r_{+}^{2}+9\alpha^2 r_+-\alpha^3)
\over (9 r_+ -\alpha)^3}\right],\\
 \mathcal{R}_+&\equiv& \mathcal{R}_+(r=r_+)
\end{eqnarray}
Therefore, at this value of the angular momentum, $L_{lsco}$, the {\it last stable circular orbit} occurs
as is shown in Fig. \ref{f1}.

In addition with the principal equation of motion (\ref{tl7}),
the remaining quadratures are written as
\begin{eqnarray}\label{t20}
  \left(\frac{dr}{dt}\right)^2 &=& \frac{\mathcal{F}\,\left(\,E-V_t\,\right)}{E}, \\\label{t21}
  \left(\frac{dr}{d\phi}\right)^2 &=& \frac{\mathcal{R}^4\,\left(\,E-V_t\,\right)}{L^2}.
\end{eqnarray}
First, from Eq. (\ref{tl8}) we note that neutral massive particles with $L = 0$
have the same behavior as in the Schwarzschild space-time studied earlier, for example, by Chandrasekhar \cite{chandra},
Wald \cite{wald}, and Schutz \cite{shutz}, among other authors.
Therefore, we only put our attention on the motion
of massive neutral particles with non-zero angular momentum.
Let us consider Eq. (\ref{t21}), and rewrite it as

\begin{eqnarray}\label{tl11}
  \left(\frac{dr}{d\phi}\right)^2&=&\frac{\mathcal{R}^2}{r\,L^2}\left[-(1-E)\,r^3+
  (\alpha(1 - E)+r_+)\,r^2\right.\nonumber\\
  &&\left.- L_{\alpha} r_+\,r+L^2\,r_+\right].
\end{eqnarray}
In order to obtain a full description of the motion of massive particles,
we separately study the two possible cases: The bound orbits ($E<1$) and the unbound orbits
($E=1$).

\subsection{The bound orbits ($E<1$)}
As is possible to see from Equation (\ref{tl8}), $V_t\rightarrow 1$
for $r\rightarrow \infty$. This means that particles with $E<1$
always have a turning point which corresponds to an {\it aphelion}
distance, and depending on
the value of the angular momentum, eventually
has others turning point, see bottom panel of Fig. \ref{f1}.

For studying the motion of particles with
this characteristic, let us to rewrite Eq. (\ref{t21}) as

\begin{equation}\label{tl11a}
  \left(\frac{dr}{d\phi}\right)^2=\frac{(1-E)}{L^2}\frac{\mathcal{R}^2\, \mathcal{P}}{r}.
\end{equation}
Here the characteristic polynomial $\mathcal{P}=\mathcal{P}(r)$ is given by

\begin{equation}\label{tl12}
\mathcal{P}=-r^3+r_b\,r^2-
L_{\alpha}\,(r_b-\alpha)\,r+L^2(r_b-\alpha),
\end{equation}
where
\begin{equation}\label{tl13}
  r_b={r_+\over 1-E}+\alpha.
\end{equation}
Therefore, depending of the nature of its roots,
we shall obtain the allowed motions for this configuration.

\subsubsection{The circular orbits}
As we have seen, particles with $L\geq L_{lsco}$ can stay in
a circular orbit at $r_{circ}$; it can be stable ($r_{circ}=r_s$) or
unstable ($r_{circ}=r_u$). The periods for one complete revolution
of these circular orbits, measured in proper time and coordinate time, are
\begin{eqnarray}
&&T_{\tau}= 2 \pi\sqrt{{{2 r_{circ}^{3}-3 r_+ r_{circ}^{2}
-\alpha( r_{circ}^{2}-2 r_+ r_{circ})} \over r_+}},\nonumber\\
&&T_{t}= 2 \pi\sqrt{ {2 r_{circ}^{3}
-\alpha r_{circ}^{2} \over r_+}}.\label{c1}
\end{eqnarray}
On the other hand, expanding the effective potential in turn
to $r=r_s$, one can write
\begin{equation}\label{c2}
V(r)=V(r_s)+V'(r_s)(r-r_s)+{1\over2}V''(r_s)(r-r_s)^2+...,
\end{equation}
where$\,'$ means derivative with respect to radial coordinate.
Obviously, in this orbits $V'(r_s)=0$, so, by defining the {\it smaller}
coordinate $x=r-r_s$, together with {\it the epicycle frequency}
$\kappa^2=1/2V''(r_s)$,  we can rewrite the above equation as
\begin{equation}\label{c3}
V(x)\approx E_s+\kappa^2\,x^2
\end{equation}
where $E_s$ is the energy of the particle at the stable circular orbit.
Also, it is easy to see that test particles satisfy the harmonic equation of motion
\begin{equation}\label{c4}
\ddot{x}=-\kappa^2\,x.
\end{equation}
In our case, the epicycle frequency is given by
\begin{equation}\label{c5}
\kappa^{2}=\kappa_{Schw}^{2} \left(\frac{1+\epsilon_1}{1+\epsilon_2}\right),
\end{equation}
where $\kappa_{Schw}$
is the epicycle frequency in the Schwarzschild case given by \cite{letelier}
\begin{equation}\label{c6}
\kappa_{Schw}^{2}= {r_+ \over r_{s}^{3}}\left({r_{s}-3r_+
 \over
2 r_{s}-3r_+}\right),
\end{equation}
and the functions that appears here are given by
\begin{eqnarray}\label{c6.1}
  \epsilon_1 &=& \frac{\alpha\,r_+}{r_s}\frac{(2 r_{s}-3r_+)}{r_s-3r_+} \\ \label{c6.2}
  \epsilon_2 &=& \frac{\alpha\,r_s\,(5r_+-3r_s)+\alpha^2(r_s-2r_+)}{2r_s-3r_+}.
\end{eqnarray}
Notice, from Eqs. (\ref{c5},  \ref{c6.1}, \ref{c6.2}),   that $\kappa \rightarrow \kappa_{Schw}$
when $\alpha\rightarrow 0$.
\subsubsection{Orbits of the first kind}
Orbits of the first kind occur when the energy lies on the range $E_s<E<E_u$,
and this case requires that $P(r)=0$ allows three real roots, all of which are positive;
and we shall write them as
\begin{equation}\label{c7}
r_{d}^{(\nu)}={r_b \over 3}+\sqrt{{\eta_2 \over 3}}
\cos \left[{1 \over 3} \arccos \left(\sqrt{{27\eta_{3}^2 \over \eta_{2}^{3}}}\right)+
{2 \pi \nu \over 3}\right],
\end{equation}
where $\nu=0, 1, 2$, and
\begin{eqnarray}\label{c8}
  \eta_2  &=& 4\left({r_b^2 \over 3}-L_{\alpha}(r_b-\alpha)\right), \label{c9} \\
  \eta_3 &=& 4\left({2r_b^3 \over 27}-
{L_{\alpha}\,\mathcal{R}_b^2 \over 3}
+ L^2(r_b-\alpha)\right),
\end{eqnarray}
where $\mathcal{R}_b\equiv \mathcal{R}(r=r_b)$. So, we can identify the aphelion distance as
$r_{d}^{(0)}=r_A$, and the perihelion distance as $r_{d}^{(2)}=r_P$,
while the third solution can be recognized as the aphelion
distance to the orbits of the second kind, $r_\textrm{a}=r_{d}^{(1)}$, see
bottom panel of Fig. \ref{f1}. In this way, we can rewrite
the characteristic polynomial (\ref{tl12}) as
\begin{equation}\label{c10}
\mathcal{P}=(r_A-r)(r-r_p)(r-r_\textrm{a}).
\end{equation}
Substituting Eq. (\ref{c10}) into Eq. (\ref{tl11a}) and then integrating,
we obtain the polar form to the first kind orbit of the neutral massive particles,
resulting
\begin{equation}\label{c11}
r(\phi)=r_A-{\alpha \over 4\wp(\kappa_1̣\,\phi; g_{2},g_{3})-\alpha_1 /3},
\end{equation}
where $\wp(x; g_2, g_3)$ is the $\wp$-Weierstra{\ss} elliptic function \cite{Weierstrass,hancock}, with
the Weierstra{\ss} invariant given by
\begin{equation}\label{c12}
g_2 ={1 \over 4}\left({\alpha_{1}^{2} \over 3}-
\beta_1\right),\quad
g_3 ={1 \over 16}\left({\alpha_{1}\beta_1 \over 3}-{2 \over 27}\alpha_{1}^{3}-\gamma_1\right),
\end{equation}
while the constants are given explicitly by
\begin{eqnarray}\nonumber
  \kappa_1&=&\frac{1}{L}\sqrt{{(1-E)(r_A-\alpha)( r_A-r_P)(r_A-r_{\textrm{a}})\over  \alpha}},\\ \nonumber
  \alpha_1&=&-\alpha\left[{1\over r_A-\alpha}+{1\over r_A-r_P}+
{1\over r_A-r_{\textrm{a}}}\right]\\ \nonumber
\beta_1&=&\alpha^{2}\left[{1\over (r_A-\alpha)( r_A-r_P)}+
{1\over (r_A-r_{\textrm{a}})( r_A-\alpha)}+\right.\\
&&+\left.{1\over (r_A-r_P) (r_A-r_{\textrm{a}})}\right],\\ \nonumber
\gamma_1&=&-\frac{(1-E)\,\alpha^2}{L^2\,\kappa_1^2}.
\end{eqnarray}
In the top panel of Fig. \ref{f2} we plot the polar trajectory (\ref{c11}),
and it shows that orbits precess between  the aphelion distance, $r_A$, and perihelion
distance, $r_P$.
Furthermore, we can determine the angle $\chi=2\phi_P$ corresponding to an oscillation, resulting in
\begin{equation}\label{c13}
\chi={2\over \kappa_1}\wp^{-1}\left[{ \alpha_1 \over 12}+{ \alpha \over 4(r_A-r_P)}\right].
\end{equation}
In section \ref{peri} we will present the  post-Newtonian approximation to obtain
some information about the parameters of the theory.
\subsubsection{Orbits of the second kind}
\begin{figure}[!h]
 \begin{center}
  \includegraphics[width=60mm]{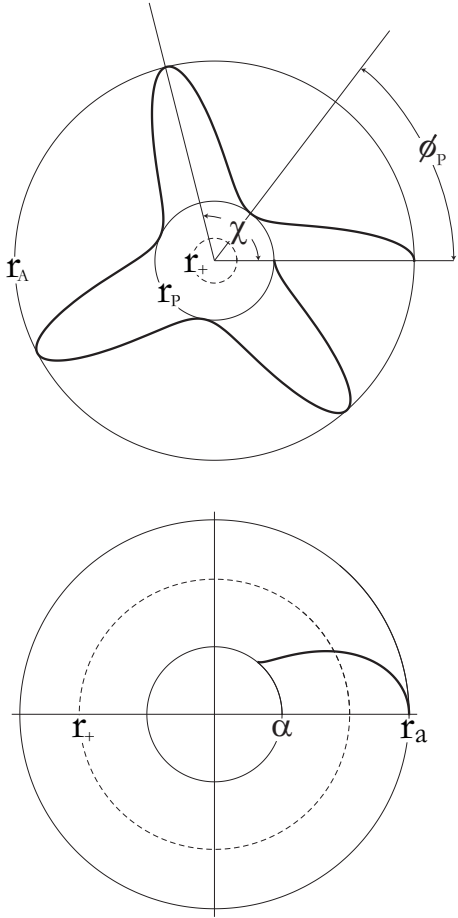}
 \end{center}
 \caption{Plot of the bound orbits for neutral massive particles. Top panel: Orbit of the first kind
 which precession between the aphelion distance, $r_A$, and the perihelion distance, $r_P$. Also, this plot shows
 the angle $\chi=2\phi_P$ corresponding to an oscillation;
 bottom panel: orbit of the second kind for which massive particles plunge from the aphelion distance, $r_{\textrm{a}}$, to the interior
 singular surface at $r=\alpha$.}
 \label{f2}
\end{figure}

As we have already explained, orbits of the
second kind have their aphelions at $r_{\textrm{a}}$
and eventually plunge to the radial distance $r=\alpha$.
Therefore, in this case we have the following
characteristic polynomial:
\begin{equation}\label{c14}
  \mathcal{P}=(r_A-r)(r_P-r)(r_{\textrm{a}}-r),
\end{equation}
so, inserting Eq. (\ref{c14}) into Eq. (\ref{tl11a}) and integrating,
we obtain
\begin{equation}\label{c15}
r(\phi)=r_{\textrm{a}}-{\alpha \over 4\wp(\kappa_2\,\phi; g_{2}, g_{3})-\alpha_2 /3}.
\end{equation}
Again, $\wp(x; g_2, g_3)$ is the $\wp$-Weierstra{\ss} elliptic function
with the Weierstra{\ss} invariants given by
\begin{equation}\label{c16}
g_2 ={1 \over 4}\left({\alpha_{2}^{2} \over 3}-
\beta_2\right),\quad
g_3 ={1 \over 16}\left({\alpha_{2}\beta_2 \over 3}-{2 \over 27}\alpha_{2}^{3}-\gamma_2\right),
\end{equation}
where the constants are given by
\begin{eqnarray}\nonumber
\kappa_2&=& \frac{1}{L}\sqrt{{[(1-E)(r_{\textrm{a}}-\alpha)( r_A-r_{\textrm{a}})(r_P-r_{\textrm{a}})]\over  \alpha }}
\\ \nonumber
\alpha_2&=&\alpha\left[{1\over r_A-r_{\textrm{a}}}+{1\over r_P-r_{\textrm{a}}}-{1\over r_{\textrm{a}}-\alpha}\right]
\\ \nonumber
\beta_2&=&\alpha^{2}\left[{1\over (r_A-r_{\textrm{a}})( r_P-r_{\textrm{a}})}-
{1\over (r_A-r_{\textrm{a}})( r_{\textrm{a}}-\alpha)}+\right.\\
&&\left.-
{1\over (r_P-r_{\textrm{a}}) (r_{\textrm{a}}-\alpha)}\right]
\\ \nonumber
\gamma_2&=&-\frac{(1-E)\,\alpha^2}{L^2\,\kappa_2^2}
\end{eqnarray}
The polar trajectory of the second kind is plotted in the bottom panel of
Fig. \ref{f2}.

\subsubsection{The critical trajectories}
\begin{figure}[!h]
 \begin{center}
  \includegraphics[width=74mm]{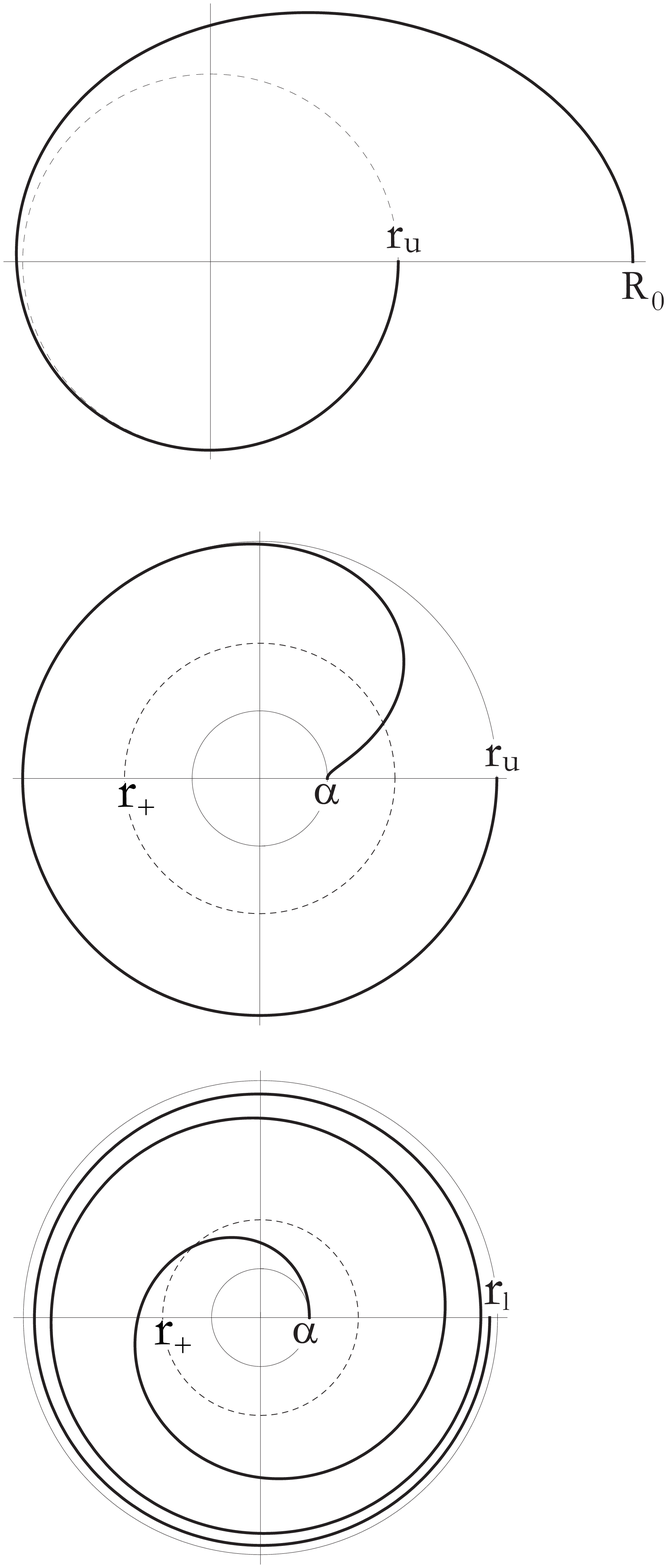}
 \end{center}
 \caption{Plot for critical bound orbits. Top panel: critical orbit of the first kind, whose trajectories followed by
 particles falling from a distance $R_0>r_u$  and then approach asymptotically to the unstable circular
 orbit at $r_u$. Here the initial condition $r(\phi=0)=R_0$ has been used
 ; middle panel: critical orbit of the second kind corresponding to the trajectories for massive particles
 that can go asymptotically to the unstable circular orbit at $r_u$ or can fall into the event horizon
 and eventually go to the singular surface at $r=\alpha$; bottom panel: last circular orbit allowed by this space-time.
 The two last graphics have been made with the initial condition $r(\phi=0)=\alpha$.}
 \label{f3}
\end{figure}

Neutral massive particles follow critical trajectories when
their energy is $E_u$ (see bottom panel of Fig. \ref{f1}), which, of course, satisfy the condition $E_u=V_{t}(r_u)$.
Considering that the motion is performed in the region $r_u<r<R_0$,
we get the characteristic polynomial as
\begin{equation}\label{c17}
  \mathcal{P}=(R_0-r)(r-r_u)^2,
\end{equation}
and thus, the polar trajectory becomes
\begin{equation}\label{c18}
r(\phi)=R_0-{( R_0-r_u)(R_0-\alpha) \over ( R_0-r_u)+(r_u-\alpha)\coth^{2}(\omega_0\, \phi)},
\end{equation}
where the constants are
\begin{equation}\label{c19}
  R_0\equiv r_{d}^{(0)}(E_u),\quad
  \omega_0={\sqrt{(1-E_u)(R_0-r_u)(r_u-\alpha)}\over  2 L},
\end{equation}
and we have plotted this motion in the top panel of Fig. \ref{f3}.

On the other hand, if the motion is performed in the region
$r_0< r<r_u$, where $r_0$ is starting distance, then the characteristic polynomial is
\begin{equation}\label{c20}
  \mathcal{P}=(R_0- r)(r_u-r)^2,
\end{equation}
in which case we obtain the following polar orbit
\begin{equation}\label{c21}
r(\phi)=\alpha+{( R_0-\alpha)(r_u-\alpha) \over (r_u-\alpha)+( R_0-r_u)\coth^{2}(\omega_0\,\phi)}.
\end{equation}
Finally, we can obtain the polar trajectory for the last circular orbit (see top panel
of Fig. \ref{f1})
\begin{equation}\label{c22}
r(\phi)=r_l-{4 L_{lsco}^{2}\,(r_l-\alpha) \over 4 L_{lsco}^{2}\,+\,(1-E_l)\,[(r_l-\alpha)\, \phi]^{2}}.
\end{equation}

\subsection{The unbound orbits  ($E\geq1$)}

The unbound orbits are those trajectories where
massive particles posses an energy $E\geq 1$. Without loss of generality,
we choose the orbits with $E=1$ for the description of this class of orbits.
Therefore, the motion equation can be written as
\begin{equation}\label{u0}
  \left(\frac{dr}{d\phi}\right)^2= \frac{\mathcal{R}^2\,\mathcal{B}}{(L_{\alpha}-\alpha)\,r},
\end{equation}
where the polynomial $\mathcal{B}=\mathcal{B}(r)$ is given by
\begin{equation}\label{u1}
  \mathcal{B}=r^2-\,L_{\alpha}\,r+L^2=(r-r_1)(r-r_2).
\end{equation}

\begin{figure}[!h]
  \includegraphics[width=80mm]{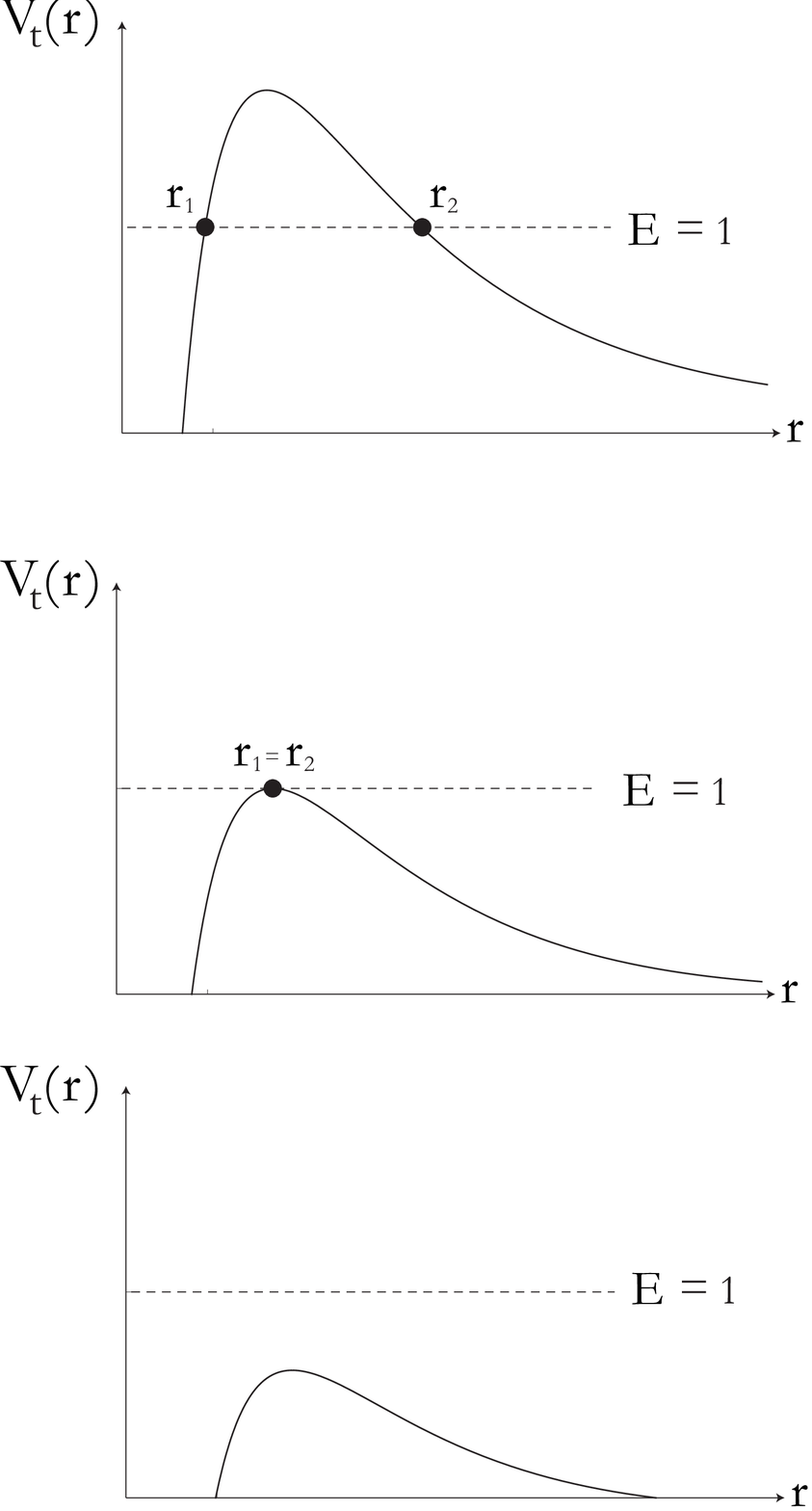}
 \caption{The disposition of the roots of the quadratic equation $\mathcal{B}=0$ for $E=1$.
 Top panel: distinct positive roots with $r_2>r_1$; middle panel: coincident  positive roots, $r_1=r_2\equiv\rho_u$;
 bottom panel: complex roots, $r_1=r_2^*$.}
 \label{f4}
\end{figure}

Since the constant term in $\mathcal{B}$ is positive, the equation $\mathcal{B}=0$ must allow
either two positive roots or a complex pair. Thus, the cases we must distinguish are those shown in Fig. \ref{f4}.
Provided that there are two roots and they are positive (distinct or coincident),
we must continue to distinguish between orbits of two kind:
orbits of the first kind restricted to the interval,
$r_2\leq r <\infty$ (which are the analogues of the
hyperbolic orbits of the Newtonian theory)
and the orbits of the second kind with $r\leq r_1$
(which are, in essence, no different from the bound
orbits of the second kind). When $r_1=r_2$, the two
kinds of orbits coalesce as they approach, asymptotically,
a common circle from opposite side by spiralling
round it an infinite number of times. Finally, when
the equation $\mathcal{B}=0$ allows a pair of complex-conjugate
roots, the resulting orbits
can be considered as belonging to imaginary eccentricities.

\subsubsection{Orbits of the first and second kinds}
Here we study the motion of test particles by considering
that two real distinct roots of the equation $\mathcal{B}=0$ are allowed.
So, recalling
\begin{equation}\label{u2}
  \rho_u=\frac{1}{2}\,L_{\alpha},\quad \textrm{and}\quad
  \Delta =\sqrt{1-\frac{4L^2}{\rho_u^2}},
\end{equation}
we can write the real (positive) roots as
\begin{equation}\label{u3}
r_1=\rho_u\,(1-\Delta), \qquad r_2=\rho_u\,(1+\Delta).
\end{equation}
So, by considering the orbits of the first kind ($r>r_2$), we
obtain the polar form of the trajectory in terms of the Weierstra{\ss} elliptic function,
which becomes
\begin{equation}\label{u4}
r(\phi)={a_1 \over3} +4\wp \left(\frac{\phi-\Omega_0}{\sqrt{L_{\alpha}-\alpha}}; g_{2},g_{3}\right),
\end{equation}
where the Weierstra{\ss} invariants read
\begin{equation}\label{u5}
  g_2 ={1 \over 4}\left[{a_{1}^{2} \over 3}-
b_1\right],\quad
g_3 ={1 \over 16}\left[-{a_{1}b_1 \over 3}+{2 \over 27}a_{1}^{3}+c_1\right],
\end{equation}
while the constants are given by
\begin{equation}\label{u6}
a_1=L_{\alpha}+ \alpha,\quad
b_1=\alpha\,L_{\alpha}+L^2,\quad
c_1=\alpha L^2.
\end{equation}
Notice that in Eq. (\ref{u4}) we have employed the initial condition
$r(0)=r_2$, which implies that
\begin{equation}\label{u61}
  \Omega_0=\wp^{-1} \left({r_2-a_1/ 3 \over 4}  ; g_{2},g_{3}\right).
\end{equation}

%%%%%%%%%%%%%%%%%%%%%%%%%%%%%%%%%%%%%%%%%%%%%
%%%%%%%%%%%%%%%%%%%%%%%%%%%%%%%%%%%%%%555
From the solution (\ref{u4}) the deviation angle, $\varphi=2\phi_{\infty}$, for this trajectory
is obtained directly, and it reads

\begin{equation}\label{u6.1}
 \varphi =2\,\sqrt{L_{\alpha}-\alpha}\,\Omega_0 .
\end{equation}

\begin{figure}[!h]
 \begin{center}
  \includegraphics[width=80mm]{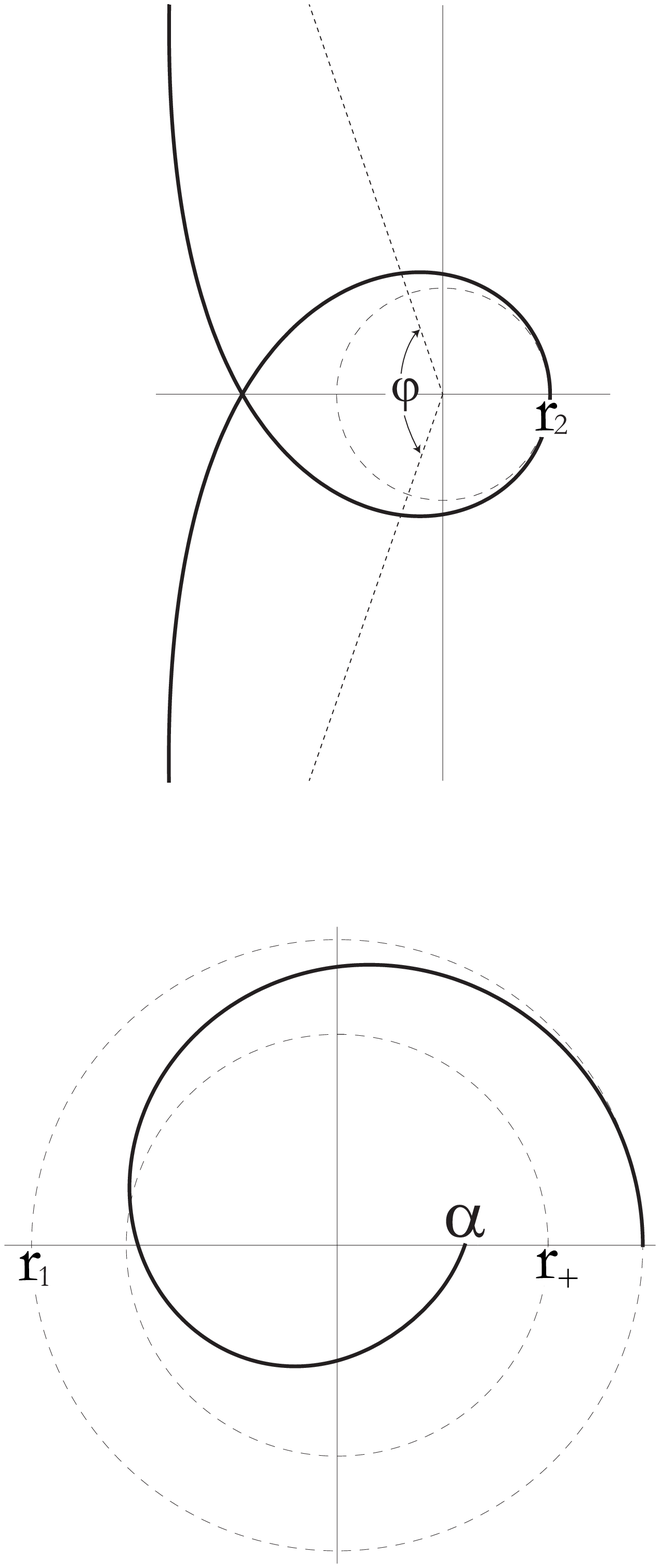}
 \end{center}
 \caption{The unbounded orbits with $E=1$. Top panel: orbits of the first kind. Particles
 coming from the spatial infinity and approach to the minimal distance $r_2$ and then going again to the spatial infinity. Also,
 the deviation angle $\varphi=2\phi_{\infty}$ is shown in this plot;
 bottom panel: orbit of the second kind for massive particles starting from the turning point at $r_1$ which
 cross the event horizon and arrive at the singular surface at $r=\alpha$.}
 \label{f5}
\end{figure}

%%%%%%%%%%%%%%%%%%%%%%%%%%%%%%%%%%%%%%%%%%%%%%
%%%%%%%%%%%%%%%%%%%%%%%%%%%%%%%%%%%%%%%%%%%%%%%
In the same way, when the orbits of the second kind is taken account ($r<r_1$),
we obtain the polar form of the trajectory, which is given by

\begin{equation}\label{u7}
r(\phi)={a_1 \over3} +4\wp \left(\frac{\Xi_0-\phi}{\sqrt{L_{\alpha}-\alpha}}; g_{2}, g_{3}\right),
\end{equation}
where the initial condition $r(0)=r_1$ yields 
\begin{equation}\label{u8}
\Xi_0=\wp^{-1} \left({r_2-a_1/ 3 \over 4}  ; g_{2},g_{3}\right),
\end{equation}
with same set of constant (\ref{u5}) and (\ref{u6}). The orbits of the first and second kind
(cf. Eqs. (\ref{u4}) and (\ref{u7})) are plotted in Fig. \ref{f5}.

\subsubsection{The critical trajectories}
\begin{figure}[!h]
 \begin{center}
  \includegraphics[width=90mm]{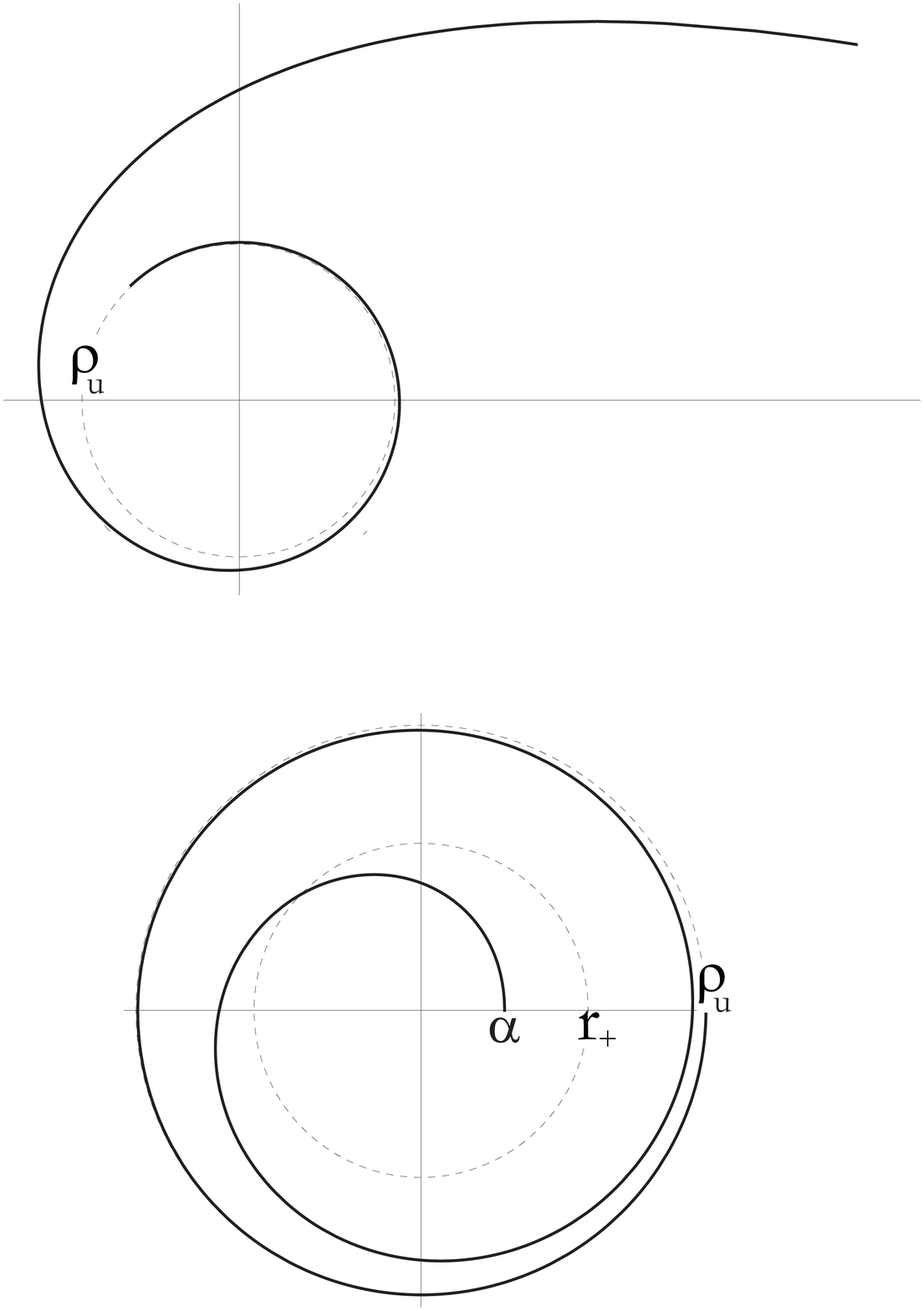}
 \end{center}
 \caption{The unbounded critical trajectories. Top panel: orbits of the first kind for massive particles
 which are coming from spatial infinity and fall asymptotically to the unstable
 circular orbit at $\rho_u$;
 bottom panel: orbits of the second kind, corresponding to the trajectories followed by neutral massive
 particles which, depending on their initial velocity, can go either to the singular surface at $r=\alpha$ or to the unstable
 circular orbit at $\rho_u$.}
 \label{f6}
\end{figure}

As we have said, the neutral massive particles can
fall asymptotically from a distance $r_i$ to an unstable circular orbit at $r_1 =r_2\equiv \rho_u$, see Fig. \ref{f4}.
Let us call  $L_u$  to  the solution of equation $E=1=V_t (\rho_u, L_u)$ , which is given by
\begin{equation}\label{u9}
L_{u}=\frac{\mathcal{R}_u}{\sqrt{s+1}},
\end{equation}
where $\mathcal{R}_u=\mathcal{R}(r=\rho_u)$ and $s=\rho_u/r_+$.
Therefore, if $\rho_u < r_i <\infty$, then the polar trajectory is given by
\begin{equation}\label{u10}
r(\phi)=\alpha +(\rho_u-\alpha)\coth^{2} \left[\frac{\sqrt{1+s^{-1}}}{2}\, \phi\right],
\end{equation}
where we have used the condition $ \phi = 0$ when $ r \rightarrow \infty $.
On the other hand, if $\alpha<r_i<\rho_u$, then the polar orbit can be written as
\begin{equation}\label{u11}
r(\phi)=\alpha +(\rho_u-\alpha)\tanh^{2} \left[\frac{\sqrt{1+s^{-1}}}{2}\, \phi\right],
\end{equation}
where, for simplicity, we have used the condition $ \phi = 0$ when $ r = \alpha $.
The critical trajectories, (\ref{u10}) and (\ref{u11}), are plotted in Fig. \ref{f6}.

\subsubsection{The last orbit}

\begin{figure}[!h]
 \begin{center}
  \includegraphics[width=50mm]{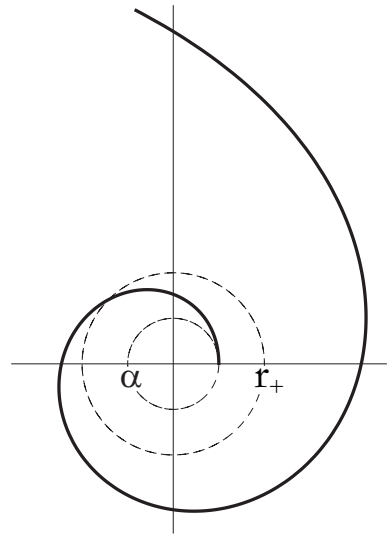}
 \end{center}
 \caption{The last orbit. These trajectories correspond to those followed by
 massive particles coming from spatial infinity which cross  inexorably the event  horizon and then arrive
 at the singular surface at $r=\alpha$.}
 \label{f7}
\end{figure}

Finally, by considering test particles with angular momentum $L_{lsco}<L<L_u$,
we can obtain the last orbit allowed in this space-time, and it correspond to
the situation depicted in the right panel of Fig.\ref{f4}. Therefore, starting
with the initial condition $\phi=0$ at $r=\alpha$, the analytic solution
to this motion is given by
\begin{equation}\label{u12}
r(\phi)={a_1 \over3} +4\wp \left(\frac{\Theta_0+\phi}{L_{\alpha}-\alpha}; g_{2}, g_{3}\right),
\end{equation}
where the Weierstra{\ss} invariants are given in Equation (\ref{u5}), and
the initial condition $r(0)=\alpha$ yields
\begin{equation}\label{u13}
\Theta_0=\wp^{-1} \left(\frac{\alpha-a_1/ 3}{4}; g_{2}, g_{3}\right) .
\end{equation}
The polar trajectory given in Eq. (\ref{u12}) is shown in Fig. \ref{f7}.

\section{Perihelion precession}\label{peri}
The following treatment, performed by Cornbleet \cite{cornbleet},
allows us to derive the formula for the advance of the perihelia of
planetary orbits. The starting point is to consider
the line element in unperturbed Lorentz coordinates
\begin{equation}\label{pp1}
  ds^2=-dt^2+dr^2+r^2(d\theta^2+\sin^2\theta d\phi^2),
\end{equation}
together with the GMGHS line element (\ref{tl1}). So, considering
only the radial and time coordinates in the binomial
approximation, the transformation gives
\begin{equation}\label{pp2}
  d\widetilde{t}\approx\left(1-\frac{r_+}{2r}\right)dt\quad \textrm{and}\quad
  d\widetilde{r}\approx\left(1+\frac{r_+}{2r}\right)dr.
\end{equation}
We will consider two elliptical orbits, one the classical Kepler orbit
in $(r, t)$ space and a GMGHS orbit in an $(\widetilde{r}, \widetilde{t})$ space.
Then in the Lorentz space $dA=\int_{0}^{\mathfrak{R}}rdr\,d\phi=\mathfrak{R}^2\,d\phi/2$, and hence
\begin{equation}\label{pp3}
  \frac{dA}{dt}=\frac{1}{2}\mathfrak{R}^2\frac{d\phi}{dt},
\end{equation}
which corresponds to Kepler's second law.
For the GMGHS case we have
\begin{equation}\label{pp4}
  d\widetilde{A}=\int_0^\mathfrak{R} \mathcal{R}\,d\widetilde{r}\,d\phi,
\end{equation}
where $d\widetilde{r}$ is given by Eq. (\ref{pp2}), and
the binomial approximation for the radial function (\ref{tl2})
is
\begin{equation}\label{pp5}
  \mathcal{R}\approx r\left(1-\frac{\alpha}{2r}\right).
\end{equation}
So, we can write (\ref{pp4}) as
\begin{eqnarray}\nonumber
  d\widetilde{A}&=&\int_0^\mathfrak{R} \left(r-\frac{\alpha}{2}\right)\left(1-\frac{r_+}{2r}\right)dr\,d\phi\\\label{pp6}
  &\approx& \frac{\mathfrak{R}^2}{2}\left(1+\frac{r_+-\alpha}{\mathfrak{R}}\right)d\phi.
\end{eqnarray}
Therefore, applying the binomial approximation wherever necessary, we obtain
\begin{eqnarray} \nonumber
  \frac{d\widetilde{A}}{d\widetilde{t}} &=& \frac{1}{2}\mathfrak{R}^2\left(1+\frac{r_+-\alpha}{\mathfrak{R}}\right)\frac{d\phi}{d\widetilde{t}} \\\nonumber
  \frac{d\widetilde{A}}{d\widetilde{t}} &=& \frac{1}{2}\mathfrak{R}^2\left(1+\frac{r_+-\alpha}{\mathfrak{R}}\right)\left(1+\frac{\alpha}{2\mathfrak{R}}\right)\frac{d\phi}{dt} \\\label{pp7}
  \frac{d\widetilde{A}}{d\widetilde{t}} &=& \frac{1}{2}\mathfrak{R}^2\left[1+\frac{3r_+}{2\mathfrak{R}}\left(1-\frac{2\alpha}{3r_+}\right) \right]\frac{d\phi}{dt}
\end{eqnarray}
So, using this increase to improve the elemental angle from
$d\phi$ to $d\widetilde{\phi}$ then for a single orbit
\begin{equation}\label{pp8}
  \int_{0}^{\Delta\widetilde{\phi}}d\widetilde{\phi}=
  \int_{0}^{\Delta\phi=2\pi}\left[1+\frac{3r_+}{2\,\mathfrak{R}}\left(1-\frac{2\alpha}{3r_+}\right) \right]d\phi.
\end{equation}
The polar form of an ellipse
is given by
\begin{equation}\label{pp9}
  \mathfrak{R}=\frac{\ell}{1+\varepsilon \cos\phi},
\end{equation}
where $\varepsilon$ is the eccentricity and $\ell$ is the
semi-latus rectum.
In this way, plugging Eq. (\ref{pp9}) into Eq. (\ref{pp8}), we obtain
\begin{eqnarray}\nonumber
  \Delta\widetilde{\phi}&=&2\pi+\frac{3r_+}{2\ell}\left(1-\frac{2\alpha}{3r_+}\right)\int_{0}^{\Delta\phi=2\pi}(1+\varepsilon \cos\phi)d\phi\\\label{pp10}
  &=&2\pi+\frac{3\pi r_+}{\ell}-\frac{2\pi\alpha}{\ell}.
\end{eqnarray}
Therefore, the perihelion advance has the standard value $\Delta\phi_{GR}=3\pi r_+/\ell$ plus (minus)
the correction term coming from the string theory $\Delta\phi_{st}=-2\pi\alpha/\ell$.

\begin{figure}[!h]
 \begin{center}
  \includegraphics[width=75mm]{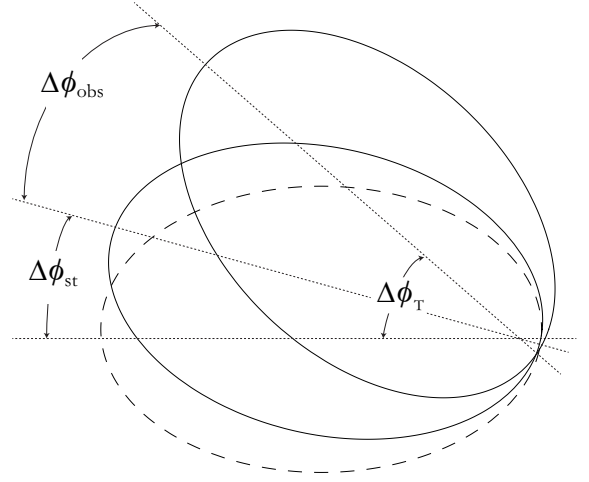}
 \end{center}
 \caption{The advance of perihelion. The heterotic correction gives an angle which increases the total effect, so
 this correction allows us to explain the discrepancy between theoretical and observational values.}
 \label{f8}
\end{figure}

\begin{table}[ht]
\caption{Sources of the precession of perihelion for Mercury}
\begin{tabular}{|c|c|}
  \hline
  \textbf{Amount} & \textbf{Cause} \\
  (arcsec/Julian century) & \\ \hline
  $\Delta\phi_{eq}=5028.83\pm 0.04$\,\cite{nasa} & \textrm{General precession in longitude} \\
  & (precession of the equinoxes)\\ \hline
  $\Delta\phi_{pl}=530$\, \cite{matzner} & \textrm{Gravitational tugs of the} \\
  & other planets \\ \hline
  $\Delta\phi_{obl}=0.0254$\, \cite{Iorio:2004ee} & \textrm{Oblateness of the Sun } \\
  & (quadrupole moment)\\ \hline
  $\Delta\phi_{GR}=42.98\pm 0.04$\, \cite{K-W} & \textrm{General Relativity} \\ \hline
  $\Delta\phi_{T}=5603.24$& \textrm{Total} \\ \hline
  $\Delta\phi_{ob}=5599.74\pm0.41$\,\cite{clemence} & \textrm{Observed} \\ \hline
  $\Delta\phi_{dis}=-3.50$ & \textrm{Discrepancy} \\
  \hline
\end{tabular}
\label{t1}
\end{table}
Aiming to explain our results, let us consider the values for Mercury given in table \ref{t1} and then
we add together the parts contributed by the general precession in longitude, the gravitational tugs
of the other planets, oblateness of the sun and the general relativity term,
we can write the total advance of perihelion as
\begin{equation}\label{pp11}
  \Delta\phi_{T}=\Delta\phi_{eq}+\Delta\phi_{pl}+\Delta\phi_{obl}+\Delta\phi_{GR},
\end{equation}
so, by comparing with the observational value, $\Delta \phi_{obs}$, we obtain a (negative)
discrepancy between both values which can be assumed as {heterotic} correction
of the advance of perihelion, i. e., we assume that $\Delta\phi_{dis}\simeq\Delta\phi_{st}$, see Fig. \ref{f8}.
Therefore, we obtain the numerical value $\alpha=0.359\,[\textrm{Km}]$ which drive us to estimate
the {\it heterotic solar charge},
$Q_{\odot}\simeq0.728\,[\textrm{Km}]= 0.493\, M_{\odot}= 8.45\times10^{19}\,[C]$.
Notice that this effect is perceived by uncharged particles,
since merely corresponds to an effect of curvature in space-time,
in the same way that photons are deflected by a massive body.
Finally, by considering that the number of electric charges is $n_{\odot}=Q_{\odot}/e\simeq 5.27\times10^{38}$,
where $e$ is the fundamental charge, we obtain the {\it heterotic solar electric density}, resulting
\begin{equation}\label{pp12}
  \rho_{\odot}=3.76\times10^{11}\,\left[\frac{{\rm electric\, charges}}{\textrm{m}^3}\right].
\end{equation}

\section{Summary}\label{summ}
In this paper we have studied the motion
of test neutral massive particle
in a Gibbons--Maeda--Garfinkle--Horowitz--Strominger
(GMGHS) black hole. The equations for the
geodesics in the Einstein frame
were exactly solved for various values of
the constants of motion of the test particles.
We note that radial motion result to be
equivalent to the Schwarzschild space-time, found
early by other authors (for example, Chandrasekhar
\cite{chandra}, Schutz \cite{shutz}, Wald \cite{wald}), so
radial massive particles cannot feel any difference with the Schwarzschild counterpart.
Moreover, the motion with non-zero angular momentum
is different, nevertheless still
presents some equivalences: there are bound and unbound orbits.
For bound orbits, we studied the circular
motion found the epicycle frequency, which reduces to
the Schwarzschild case when $\alpha\rightarrow 0$
\cite{letelier}. Also, orbits of the first and second kind
are described by means of the $\wp$-Weierstra{\ss}
elliptic function, and finally,
the two class of critical trajectories,
together with the last stable circular orbit,
are obtained in terms of elemental functions.

On the other hand, the unbound orbits
are studied for a specific value of the energy
$E=1$ since there are no change in the
physics when the case $E>1$ is considered.
Thus, again the polar form of orbits of the first and second kind
was obtained in terms of the $\wp$-Weierstra{\ss}
elliptic function, while the critical orbits
were obtained in terms of elemental functions.
In order to complete
the geodesic structure of the GMGHS space-time,
we obtain the last trajectory
that fall into the horizon which also
is obtained in terms of the $\wp$-Weierstra{\ss}
elliptic function.

Additionally, we use a method developed by Cornbleet \cite{cornbleet}
to obtain the correction to the advance of the perihelion in this
space-time which, by comparing with the available Mercury's data, allows us
to estimate the {\it heterotic solar charge}, resulting
$Q_{\odot}\simeq 0.728\,[\textrm{Km}]= 0.493\, M_{\odot}$.

Finally, in addition with the work on null geodesics developed by Fernando \cite{fernando},
this contribution completes the full geodesic structure of the GMGHS space-time.

%%%%%%%%%%%%%%%%%%%%%%%%%%%%%%%%%%%%%%%%%%%%%%%%%%%%%%%%%%%%%
%%%%%%%%%%%%%%%%%%%%%%%%%%%%%5
\begin{acknowledgments}
The authors thank to Nikolaus \& Katia Vogt, Michel Cur\'e and Radostin Kurtev for valuable helps.
MO thanks to PUCV and DFA-UV. JRV is supported by
FONDECYT grant $\textrm{N}^{\textrm{o}}\, 11130695$.
\end{acknowledgments}

%%%%%%%%%%%%%%%%%%%%%%%%%%%%%%%%%


\begin{thebibliography}{9}
\bibitem{gibbons88}
Gibbons G. W. and Maeda K.:
Black holes and membranes in higher dimensional theories with dilaton fields.
Nucl. Phys. B \textbf{298}, 741 (1988).

\bibitem{garfinkle91}
Garfinkle D., Horowitz G. T. and Strominger A.:
Charged black holes in string theory.
Phys. Rev. {\bf D} 43, 3140 (1991).

\bibitem{sen92}
Sen A.:
Rotating Charged Black Hole Solution in Heterotic String Theory.
Phys. Rev. Lett. \textbf{69}, 7 (1992)
[arXiv: 9204046].


\bibitem{sen95}
Sen A.:
Black Hole Solutions in Heterotic String Theory on a Torus.
Nucl. Phys. B \textbf{440}, 421-440 (1995) [arXiv: 9411187].

\bibitem{hassan}
Hassan S. F. and Sen A.:
Twisting Classical Solutions in Heterotic String Theory.
Nucl. Phys. B \textbf{375}, 103-118 (1992) [arXiv: 9109038].

\bibitem{khuri}
Khuri R. R.;
Solitons, Black Holes and Duality in String Theory.
Nucl. Phys. B (Proc. Suppl.) \textbf{61 A}, 99-110 (1998) [arXiv: 9704110].

\bibitem{gao05}
Gao C. J. and Zhang S. N.:
Topological Black Holes in Dilaton Gravity Theory.
Phys. Lett. B \textbf{612}, 127-136 (2005).

\bibitem{hackmann}
Hackmann E., Kagramanova V., Kunz J. and
Lammerzahl C.:
Analytic solutions of the geodesic equation in higher dimensional static spherically symmetric space-times.
Phys. Rev. D {\bf 78}, 124018 (2008)
[arXiv:0812.2428].

\bibitem{Hledik}
Stuchl\'{\i}k Z. and Hled\'{\i}k S.:
Properties of the Reissner-Nordstr\"{o}m space-times with a nonzero cosmological constant.
Acta Phys. Slov. \textbf{52}, 363 (2002) [arXiv: 0803.2685].



\bibitem{cosv}
Villanueva J. R., Saavedra J., Olivares M. and Cruz N.:
Photons motion in charged Anti-de Sitter black holes.
Astrophys. Space Sci. \textbf{344}, 437-446  (2013).

\bibitem{monin}
Olivares M., Saavedra J., Leiva C. and Villanueva J.R.:
Motion of charged particles on the Reissner-Nordstr\"{o}m (Anti)-de Sitter black hole Space-time.
Mod. Phys. Lett. A {\bf 26}, 2923 (2011).
[arXiv: 1101.0748]

\bibitem{calvani}
Stuchl\'ik S. and Calvani M.:
Null Geodesics in black hole metrics with non-zero cosmological constant.
Gen. Rel. Grav. {\bf 23}, 507 (1991).

\bibitem{pugliese}
Pugliese D., Quevedo H. and Ruffini R.:
Circular motion of neutral test particles in Reissner-Nordstr\"{o}m space-time.
Phys. Rev. {\bf D} 83, 024021 (2011)
[arXiv: 1012.5411].

\bibitem{maki94}
Maki T. and Shiraishi K.:
Motion of test particles around a charged dilatonic black hole.
Class. Quantum Grav. {\bf 11}, 227 (1994).

\bibitem{fernando}
Fernando S.:
Null Geodesics of Charged Black Holes in String Theory.
Phys. Rev. {\bf D} 85, 024033  (2012) [arXiv: 1109.0254].

\bibitem{bhadra}
Bhadra A.:
Gravitational lensing by a charged black hole of string theory.
Phys. Rev. {\bf D} 67, 103009 (2003) [arXiv: 0306016].

\bibitem{pradhan1}
Pradhan P. P.:
ISCOs in Extremal Gibbons-Maeda-Garfinkle-Horowitz-Strominger Black holes.
(2012) [arXiv: 1210.0221].

\bibitem{choi}
Choi J., Kim Y. and Park Y.:
The geodesic motion near hypersurfaces in the warped products space-time.
(2013) [arXiv: 1306.3020].



\bibitem{chandra}
Chandrasekhar S.:
The Mathematical Theory of Black Holes.
Oxford University Press, New York (1983).

\bibitem{COV}
Cruz N., Olivares M. and Villanueva J. R.:
The geodesic structure of the Schwarzschild anti-de Sitter Black Hole.
Class. Quantum Grav. \textbf{22}, 1167-1190 (2005)
[arXiv: 0408016].

\bibitem{germancito}
Olivares M., Rojas G., V\'asquez Y. and Villanueva J. R.:
Particles motion on topological Lifshitz black holes in 3+1 dimensions.
Astrophys. Space Sci. {\bf 347}, 83-89 (2013)
[arXiv: 1304.4297].

\bibitem{weyl}
Villanueva J. R. and Olivares M.:
On the null trajectories in conformal Weyl gravity.
JCAP {\bf 1306}, 040 (2013)
[arXiv: 1305.3922].

\bibitem{cov21}
Cruz N., Olivares M. and Villanueva J. R.:
Geodesic Structure of the Lifshitz black hole in 2+1 dimensions,
Eur. Phys. J. C {\bf73}, 2485 (2013)
[arXiv: 1305.2133].

\bibitem{vv}
Villanueva J. R. and V\'asquez Y.:
About the coordinate time for photons in Lifshitz space-times.
Eur. Phys. J. C {\bf 73}, 2587 (2013)
[arXiv: 1309.4417].

\bibitem{valeria08}
Kagramanova V., Kunz J. and L\"{a}mmerzahl C.:
Orbits in the field of a gravitating magnetic monopole.
Gen. Rel. Grav. {\bf40}, 1249  (2008)
[arXiv: 0708.1747].

\bibitem{evita0}
Hackmann E., Kagramanova  V., Kunz  J. and  L\"{a}mmerzahl C.:
Analytic solutions of the geodesic equation in axially symmetric space-times.
EPL {\bf88}, 30008 (2009)
[arXiv: 0911.1634].

\bibitem{evita}
Hackmann E, Hartmann B., L\"ammerzahl C. and Sirimachan P.:
The complete set of solutions of the geodesic equations in the space-time of a Schwarzschild black hole pierced by a cosmic string.
Phys. Rev. D {\bf 81},  064016 (2010) [arXiv: 0912.2327].

\bibitem{betti}
Hartmann B. and Sirimachan P.:
Geodesic motion in the space-time of a cosmic string.
JHEP {\bf 1008}, 110 (2010) [arXiv: 1007.0863].

\bibitem{chen}
Chen J. and Wang Y.:
Timelike Geodesic Motion in Ho$\check{r}$ava-Lifshitz Spacetime.
Int. J. Mod. Phys. A {\bf25}, 1439 (2010) [arXiv: 0905.2786].

\bibitem{zhou}
Zhou S., Chen J. and Wang Y.:
Geodesic Structure of Test Particle in Bardeen Spacetime.
Int. J. Mod. Phys. D {\bf21} 9, 1250077 (2012) [arXiv: 1112.5909].

\bibitem{sultana}
Sultana J., Kazanas D. and Said J. L.:
Conformal Weyl Gravity and perihelion precession.
Phys. Rev. D {\bf86}, 084008 (2012).

\bibitem{halisoy}
Halilsoy M., Gurtug O. and Habib Mazharimousavi S.:
Rindler modified Schwarzschild geodesics.
Gen. Rel. Grav. {\bf45} 11, 2363 (2013).

\bibitem{letelier}
Ramos-Caro J, Pedraza J. and Letelier P.:
Motion around a Monopole + Ring system: I. Stability of Equatorial Circular Orbits vs Regularity of Three-dimensional Motion.
MNRAS {\bf414}, 3105-3116 (2011) [arXiv: 1103.4616]

\bibitem{wald}
Wald R.M.:
General relativity.
The University Chicago Press, Chicago (1984).

\bibitem{shutz}
Schutz B.:
A First Course in General Relativity.
Cambridge University Press, New York (2009).


\bibitem{Weierstrass}
Weierstra{\ss} K.:
Zur Theorie der Abelschen Functionen.
Crelle's J. Math. {\bf 47} 289 (1854).

\bibitem{hancock}
Hancock H.:
Elliptic Integrals.
John Wiley \& Sons, New York (1917).
%%%%%%%%%%%%%%%%%%%%%%%%%%%%%%%%%%%%%%%%%%%%%%%%%%%%%%%%%%%%%%%%%%


\bibitem{cornbleet}
Cornbleet S.:
Elementary derivation of the advance of the perihelion of a planetary orbit.
Am. J. Phys. \textbf{61} (7), 650 - 651 (1993).

\bibitem{nasa}
 NASA Jet Propulsion Laboratory, http://ssd.jpl.nasa.gov/?constants

\bibitem{matzner}
Matzner R. A.: {\it Dictionary of geophysics, astrophysics, and astronomy}, CRC Press. p. 356. ISBN 0849328918 (2001).


\bibitem{Iorio:2004ee}
Iorio L.:
On the possibility of measuring the solar oblateness and some relativistic effects from planetary ranging.
Astron.\ Astrophys.\  {\bf 433}, 385 (2005)
[arXiv:0406041].

\bibitem{K-W}
Kraniotis G. V. and Whitehouse S. B.:
Compact calculation of the Perihelion Precession of Mercury in General Relativity, the Cosmological Constant and Jacobi's Inversion problem.
Class. Quantum Grav. \textbf{20}, 4817-4835 (2003) [arXiv: 0305181].

\bibitem{clemence}
Clemence, G. M.:
The Relativity Effect in Planetary Motions.
Rev. Mod. Phys. {\bf19} 4, 361-364 (1947).

\end{thebibliography}
\end{document}